\documentclass[10pt]{article}
\usepackage{amsmath}
\usepackage{amssymb}
\usepackage[dvips]{graphicx}
\usepackage{subfig}
\usepackage{lipsum}
\usepackage{textcomp}
\usepackage{upgreek}

\usepackage{cite}

\usepackage{color} 

\usepackage{setspace} 
\usepackage[labelfont=bf,labelsep=period,justification=raggedright]{caption}

\makeatletter
\renewcommand{\@biblabel}[1]{\quad#1.}
\makeatother

\date{}

\begin{document}

\begin{flushleft}
{\Large
\textbf{Ligand-dependent opening of the multiple AMPA receptor conductance states: a concerted model}
}
\\
Ranjita Dutta Roy$^{1,2,3}$, 
Christian Rosenmund$^{2}$, 
Stuart J Edelstein$^{4}$
Nicolas Le Nov\`ere $^{3,4,\ast}$
\\
\bf{1} Department of Medicine Solna, Karolinska Insitutet, 171 76 Stockholm, Sweden
\\
\bf{2} NWFZ, Charite Universitatsmedizin, 101 17 Berlin, Germany
\\
\bf{3} European Bioinformatics Institute (EMBL-EBI), Wellcome Trust Genome Campus, Hinxton, Cambridgeshire CB10 1SD , UK
\\
\bf{4} The Babraham Institute, Babraham, Cambridgeshire CB22 3AT, UK
$\ast$ E-mail: Corresponding lenov@babraham.ac.uk
\end{flushleft}

\section*{Abstract}

Modulation of the properties of AMPA receptors at the post-synaptic membrane is one of the main suggested mechanisms underlying fast synaptic transmission in the central nervous system of vertebrates. Electrophysiological recordings of single channels stimulated with agonists showed that both recombinant and native AMPA receptors visit multiple conductance states in an agonist concentration dependent manner. We propose an allosteric model of the multiple conductance states based on concerted conformational transitions of the four subunits, as an iris diaphragm. Our model predicts that the thermodynamic behaviour of the conductance states upon full and partial agonist stimulations can be described with increased affinity of receptors as they progress to higher conductance states. The model also predicts the existence of AMPA receptors in non-liganded conductive substates. However, the probability of spontaneous openings decreases with increasing conductances. Finally, we predict that the large conductance states are stabilized within the rise phase of a whole-cell EPSC in glutamatergic hippocampal neurons. Our model provides a mechanistic link between ligand concentration and conductance states that can explain thermodynamic and kinetic features of AMPA receptor gating.

\section*{Introduction}

One of the key mechanisms responsible for synaptic transmission is the activation of $\alpha$-amino-3-hydroxy-5-methyl-4-isoxazolepropionic acid receptors (AMPARs) at the post-synaptic membrane. AMPARs are ligand-gated ion channels that mediate fast excitatory synaptic transmission. They are concentrated at the post-synaptic density where their opening is allosterically regulated by glutamate. AMPARs are tetrameric receptors consisting of various combinations of the four classes of subunits GluA1-4 (GluR1-4) \cite{Traynelis2010}. Functional heteromeric receptors can also be formed with members of the kainate receptor family \cite{Rosenmund1998}.
AMPARs can be found in three functional states: basal (closed), active (open) and desensitized (closed). Several models of the transitions between those states have been described previously \cite{Jonas1993,Diamond1997,Raghavachari2004a}. In the absence of ligand the receptors are normally found in the basal state where the ion pore is closed. Glutamate binding to the clam shell-shaped agonist binding domain concomitant with closure of the two lobes D1 and D2 facilitates opening of the ion pore \cite{Stern-Bach2004a}. There is however evidence that ligand-gated ion channels can be present in an active state also in the absence of ligand \cite{Jackson1984, Turecek1997, Bocquet2009a}. When the ion channel is open, cations flow between the synaptic cleft and the post-synaptic cytoplasm. 

Most single-channel studies showed that AMPA receptors could appear in either three or four open states. The frequency of the population of molecules in the different conductance states depends on both the agonist concentration and the nature of the agonist. These features have been seen in both recombinant and native receptors, and in the absence of desensitization in recombinant receptors \cite{Rosenmund1998,Jin2003,Gebhardt2006a}. The effect of different agonist structures such as quisqualate, AMPA, glutamate and willardiines has been investigated, the latter of which are partial agonists and thus give rise to lower activation levels than full agonists \cite{Jin2003, Gouaux2004, Hansen2007a}. Agonist binding and unbinding can occur on both closed and open states \cite{Armstrong2000a}. Increased agonist concentration has consistently been linked to a shift towards higher conductance states \cite{Rosenmund1998, Gebhardt2006a, Smith2000, Smith2000d, Cull-Candy1987a, Jin2003}. It was therefore proposed that the AMPA receptor visits successive conductance states in an agonist-dependent manner, consistent with successive transitions of the receptor subunits \cite{Rosenmund1998}.

Here, we propose instead that these conductance states could reflect progressive concerted openings, akin to discrete positions of an iris diaphragm. We formulated a concerted transition model of the multiple conductance states and then simulated the thermodynamic and kinetic behavior of the receptor conductance states upon ligand binding. In particular, we examined whether the effect of full and partial agonists on the conductance of single-channels can be explained by visiting the conductance states sequentially due to increased agonist affinity. We also explored whether the time course of the rise phase of a synaptic event in a glutamatergic hippocampal neuron can be explained by the conductance in the population of AMPA receptors using this mechanism. Finally, a thermodynamic concerted MWC model would also predict the existence of spontaneously open states of the ion pore.    
        
\section*{Results}

\subsection*{A mechanistic model of the multiple conductance states of the AMPA receptor}

\begin{figure}[!h]
\begin{center}
\includegraphics[scale=1]{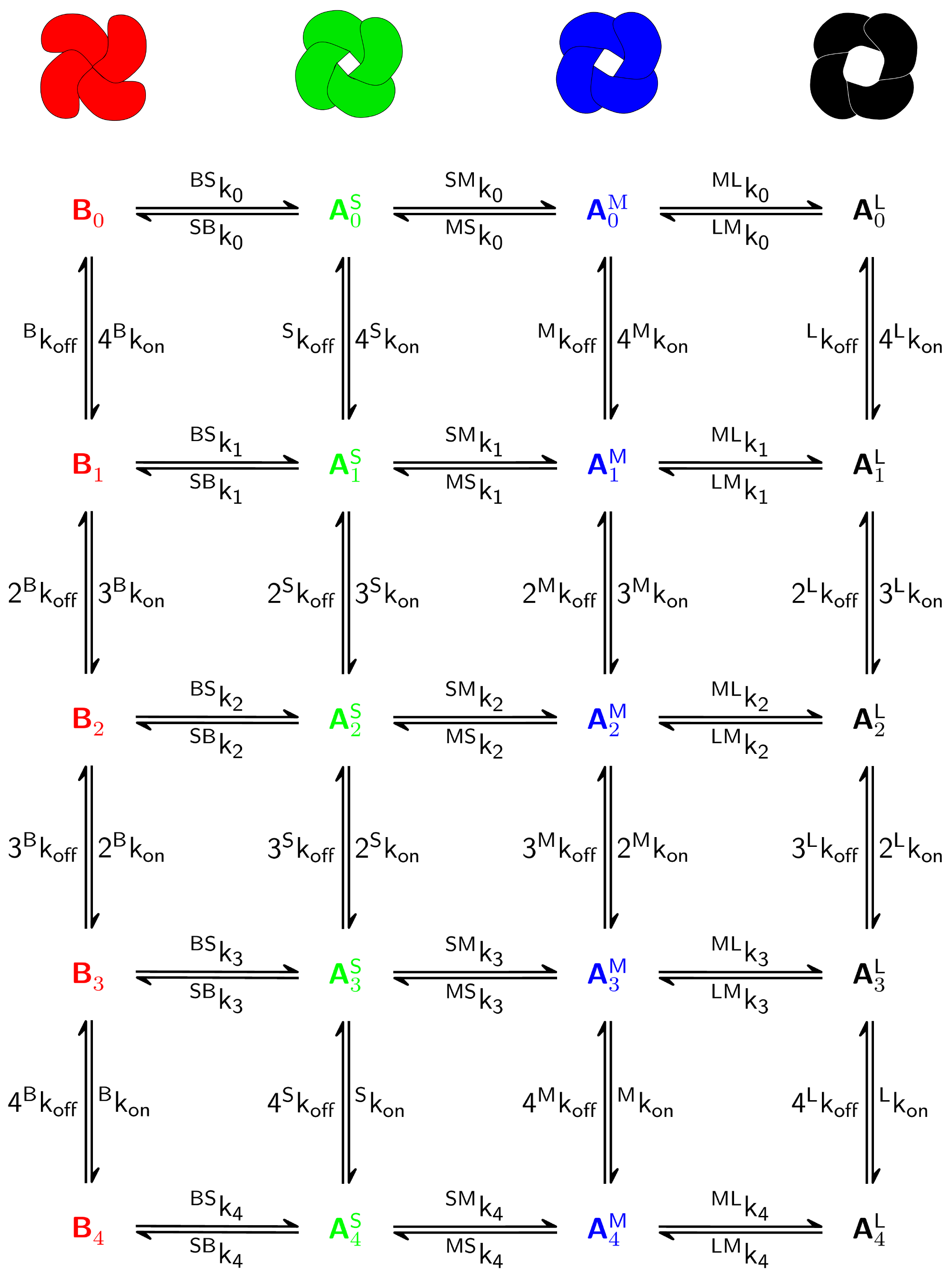}
\caption{\textbf{Kinetic scheme of the AMPAR conductance states.} The receptor can appear in either a basal (B), small (S), medium (M) or large (L) conductance state as shown on the horizontal axis. Receptors can be in any state with any binding site occupancy.}
\label{Kineticscheme}
\end{center}
\end{figure}

In order to describe the agonist-driven transitions between conductance states in the AMPA receptor, we used a model following the Monod Wyman Changeux framework \cite{Monod1965} (Supporting Equations). The receptor subunits are in equilibrium between conformations, driven by thermodynamic equilibrium. They interconvert between conformational states in a concerted manner, that is all four subunits change simultaneously. The receptor adopts distinct conformations with probabilities that depend on the extent of ligand binding. AMPA receptor molecules can be either in the basal (B) or the active small (A\textsuperscript{S}), medium (A\textsuperscript{M}) or large (A\textsuperscript{L}) states with respect to conductance as seen in Figure \ref{Kineticscheme}. Our model is parametrized with data from receptors that do not desensitize \cite{Rosenmund1998, Smith2000}. The transition between the different conformations is characterized by the state-specific allosteric constants $L_S$, $L_M$ and $L_L$, which are the ratios between the B state and each of the A states in the non-liganded form, $L=[B_0]/[A_0]$. In the absence of agonist the receptor is more likely to be found in a basal state than in a conductive state (otherwise channel opening would not be agonist-dependent) and hence all L constants are greater than one.

Ligands bind the conformations with different affinities, which explains their effect on the conformational transitions. The ratio between the state-specific dissociation constants ($K_A$ and $K_B$) of two successive liganded states is described by the state-specific $c$ parameters $c_S$, $c_M$ and $c_L$. $c$ is invariably less than one for agonists, as the active states have higher affinities for the ligand. Thus, at very low ligand concentrations, we expect most receptor molecules to be in the basal state, but more and more receptors shift to the active states as the ligand concentration is increased, since the active states are increasingly stabilized by ligand. The MWC framework distinguishes between two different functions to characterize the system's behavior. The saturation function is the fraction of ligand molecules bound to the receptor as a function of the concentration of the ligand. This function usually takes a sigmoidal shape due to the cooperativity between subunits \cite{Monod1965}. The equation for the saturation function of the AMPA receptor with three active states and a basal state is presented in supporting information S1. The state functions describe the fractions of molecules in the various conformations. 

\subsection*{Full and partial agonist effects explained by conductance dependent ligand affinity}

\begin{figure}[!h]
\begin{center}
\includegraphics[scale=1]{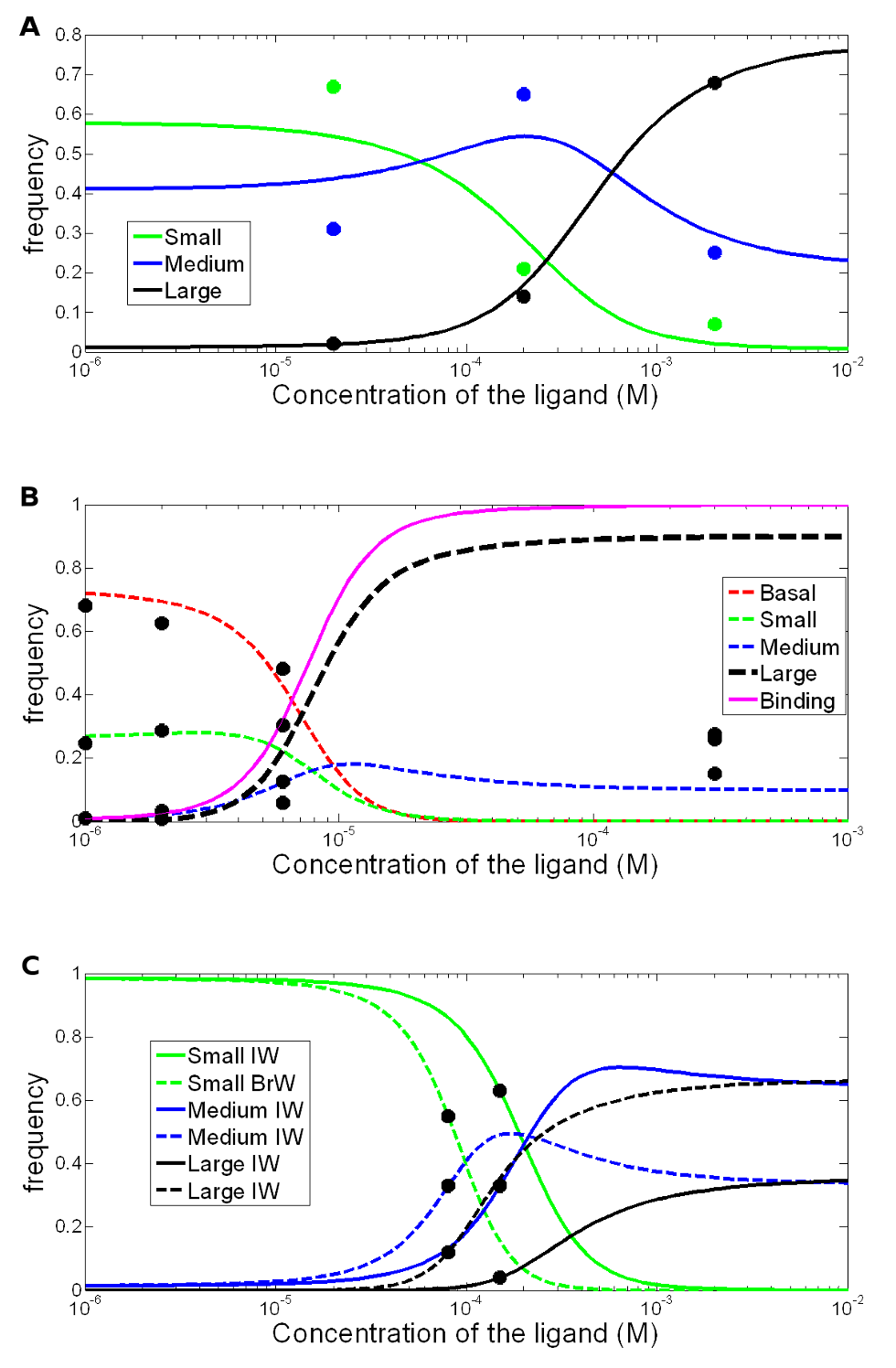}
\caption{\textbf{Effects of full and partial agonists.} (\textbf{A}) shows the small, medium and large conductance states upon stimulation with glutamate where the large state is stabilized (Supporting model S1). The relative frequency of the small conductance state (green line) is 0.6 at a ligand concentration of 1~\textmu{}M and decreases when the  ligand concentration is increased, whereas the medium conductance state (blue line) reaches its peak at a concentration above 0.1~\textmu{}M and most receptors are found in the large conductance state (black line) at 10~\textmu{}M. The dots represent experimental data \cite{Smith2000}. (\textbf{B}) stabilization of GluA3/GluK2 receptor large conductance state upon stimulation with quisqualate \cite{Rosenmund1998} (Supporting model S2). The relative frequencies of the basal state (red line) and small state (green line) of the receptor are 0.7 and 0.25 at 1~\textmu{}M and decrease when ligand concentration is increased, whereas the medium (blue) and large (black) conductance states increase and reach 0.1 and 0.9 respectively at 1~mM. The line in magenta shows the saturation function. (\textbf{C}) stabilization of GluR2 homomeric receptors intermediate conductance state upon stimulation with large willardiines (Supporting model S3). The relative frequencies of the small state (red line) is decreased and the medium (green) and large (black) states are increased when the ligand concentration increases at stimulation with both BrW and IW. At a ligand concentration of 10~mM the relative frequency of the medium state was 0.65 and 0.35 at stimulation with IW (dashed) and BrW (solid) respectively \cite{Jin2003}.}
\label{thermodynamic}
\end{center}
\end{figure}

In order to investigate whether our model could explain the behavior of native AMPA receptors, we used previously published outside-out patch clamp data. 
The data were generated from a set of native receptors in developing granule cells in the molecular layer of acute cerebellar slices stimulated by a range of glutamate concentrations in the presence of cyclothiazide to avoid desensitization \cite{Smith2000}. 
All receptors in this data set showed a tendency to dwell longer in their large conductance states with increased agonist concentrations. To investigate whether the behavior of native AMPA receptors stimulated with 20, 200 and 2000~\textmu{}M glutamate could be satisfactorily represented by our MWC model, we optimized the dissociation parameters and the ratios between the non-liganded states. The model was able to reproduce the thermodynamic profiles of the single-channels with a Pearson's correlation coefficient of $R=0.9521245$ and $p-value=7.543 \times 10^{-5}$ as seen in panel A of Figure \ref{thermodynamic}. 

A drawback of native data sets is the uncontrolled heterogeneity in subunit composition. To test our model on a more reliable data set with respect to the conductance levels, we used data from recombinant receptors in a stable transfection system, namely GluK2(GluR6)/GluA3flip receptors expressed in HEK cells \cite{Rosenmund1998}. It should be noted that GluK2 is a kainate receptor subunit, but this chimera had been used in order to avoid desensitization \cite{Stern-Bach1994} and similar transition behavior has been seen between conductance states in homomeric GluA3flip receptors \cite{Gouaux2004}. The data comes in the form of relative frequencies of conductances with different concentration of ligands. In order to deconvolute the frequencies of each subconductance state, we wrote an algorithm in Matlab (see Supplemental Materials). We focused the parameter optimization on the data obtained for 1, 2 and 6~\textmu{}M concentrations of quisqualate. Over this range of concentrations, the model fitted the data with a Pearson's correlation coefficient of $R=0.588238$ with $p-value=0.01654$ (see Figure~\ref{thermodynamic} panel B). The only moderately successful correlation between the data and the model in this case reflects the fact that the model could not be fitted well with the data at the highest glutamate concentration. For both native and recombinant receptors, the model predicted that AMPA receptors can visit all conductance states in a non-liganded form. However, with the allosteric constants in Table~\ref{table1}, we see that for native receptors, the frequency of the spontaneously open form is $10^6$ times more common at basal state compared to the large conductance state, hence, the spontaneous open form is decreased with increasing conductance. Table~\ref{table1} also lists the dissociation constants, which show that the affinity for the different states increased with the conductance state. 

Since our data was able to explain the behavior of receptors stimulated with full agonists reasonably well, we next investigated whether we could also explain why partial agonists shift receptors to an active state less efficiently than full agonists. For this we used data obtained on GluR2 homomeric receptors with the L483Y desensitizing mutation expressed in Xenopus laevis oocytes, that were stimulated by 5-substituted willardiines \cite{Gebhardt2006a}. Since the system was underdetermined, we assumed that non-liganded states were found in the same ratios as for GluA3/GluK2 receptors and performed only a parameter optimization of the dissociation constants. As seen in the panel C of Figure \ref{thermodynamic} and Table \ref{table2}, the smaller partial agonist BrW stabilized the large active state, whereas the larger IW stabilized the medium active state. It is known from other allosteric models that the effect of partial agonists can be explained by stabilization of a different state compared to full agonists \cite{Changeux1998b}. The effect of BrW can be explained by the dissociation constants, BrW having a higher affinity for the large conductance state and lower affinities for the small and medium states as seen in Table 2. The smaller partial agonist activated the receptor more efficiently than the larger partial agonist. Hence, with the presence of spontaneously active states and increased ligand affinity with conductance level, we can explain why full agonists stabilize the large conductance state and partial agonists stabilize intermediary conductance states of AMPA receptors. 

\subsection*{Stabilization of the large conductance state during the rise phase of a synaptic event}

\begin{figure}[!h]
\begin{center}
\includegraphics[scale=1]{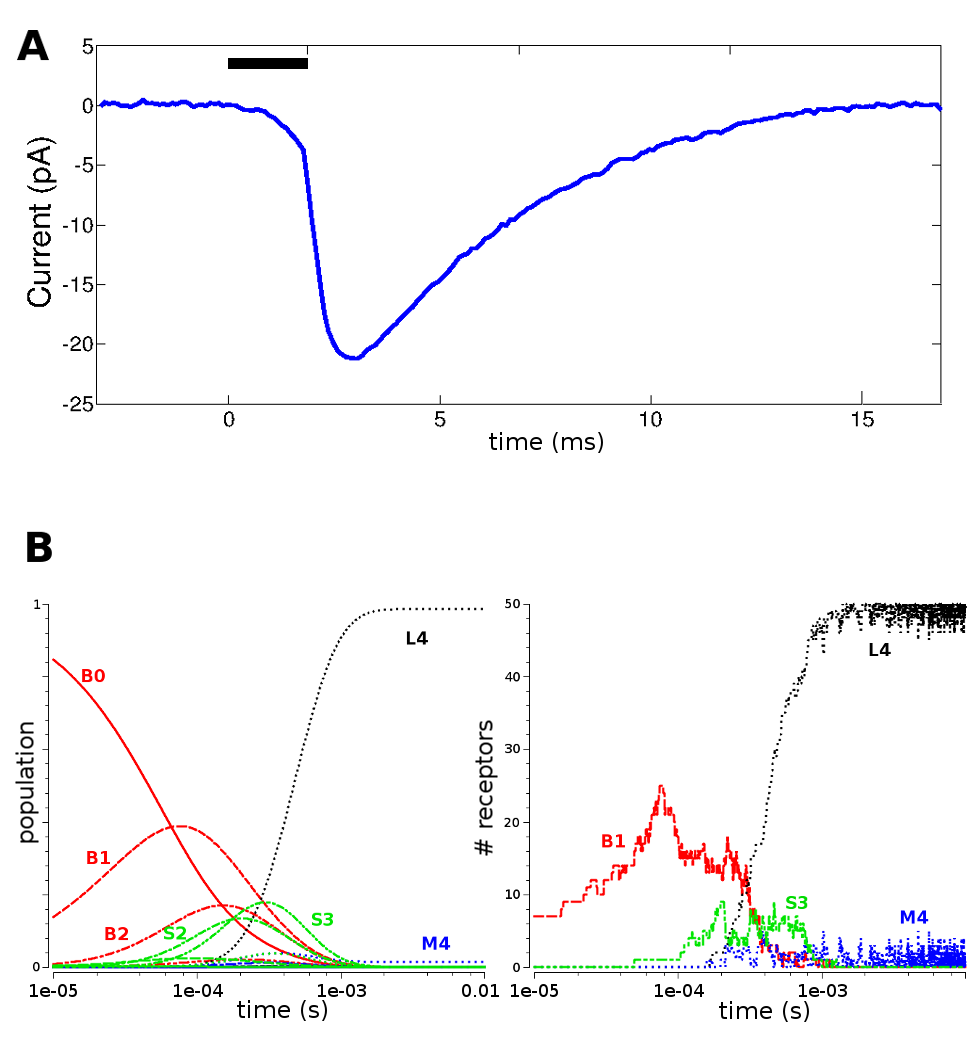}
\caption{\textbf{Kinetic behaviour of synaptic AMPARs.} (\textbf{A}) The blue trace shows the average synaptic current, which reaches its peak within fractions of ms (n=7). The black bar represents the depolarisation of the pre-synapic terminal. (\textbf{B}) Kinetics of the subconductance states of an AMPA receptor population (Supporting model S4). Left plot, deterministic simulation of a population of GluA3/GluK2 receptors by 1~\textmu{}M of agonist. Right plot, stochastic simulation of a population of 50 receptors. Only the most populated states are represented for sake of clarity.} 
\label{current}
\end{center}
\end{figure}

We next investigated whether the maximum conductance state could be stabilized within the rise time of a synaptic event. Upon release in vivo, neurotransmitters are spread and transported away from the synaptic cleft within microseconds, and the level of glutamate in the cleft can reach millimolar concentrations \cite{Lisman2007a, Clements1992a, Diamond1997}. In order to estimate the rise time of a population of receptors in a whole hippocampal neuron we averaged spontaneous whole-cell synaptic currents from wild-type autaptic hippocampal neurons as seen in Figure \ref{current}. The average $20-80$ \% rise time was $0.53$ ms $\pm$ $0.13$ ms (n=7).   

Using the kinetic scheme proposed in Figure \ref{Kineticscheme}, we formulated a kinetic model (see Supporting Information S1). The forward ligand binding rates were taken from earlier kinetic studies while the off rates were derived from the thermodynamic parameters of GluA3/GluK2 receptors (see Materials and Methods). On rates were assumed to be identical for all states. Differences in affinity were therefore determined solely by off rates. 

\begin{figure}[!h]
\begin{center}
\includegraphics[scale=1]{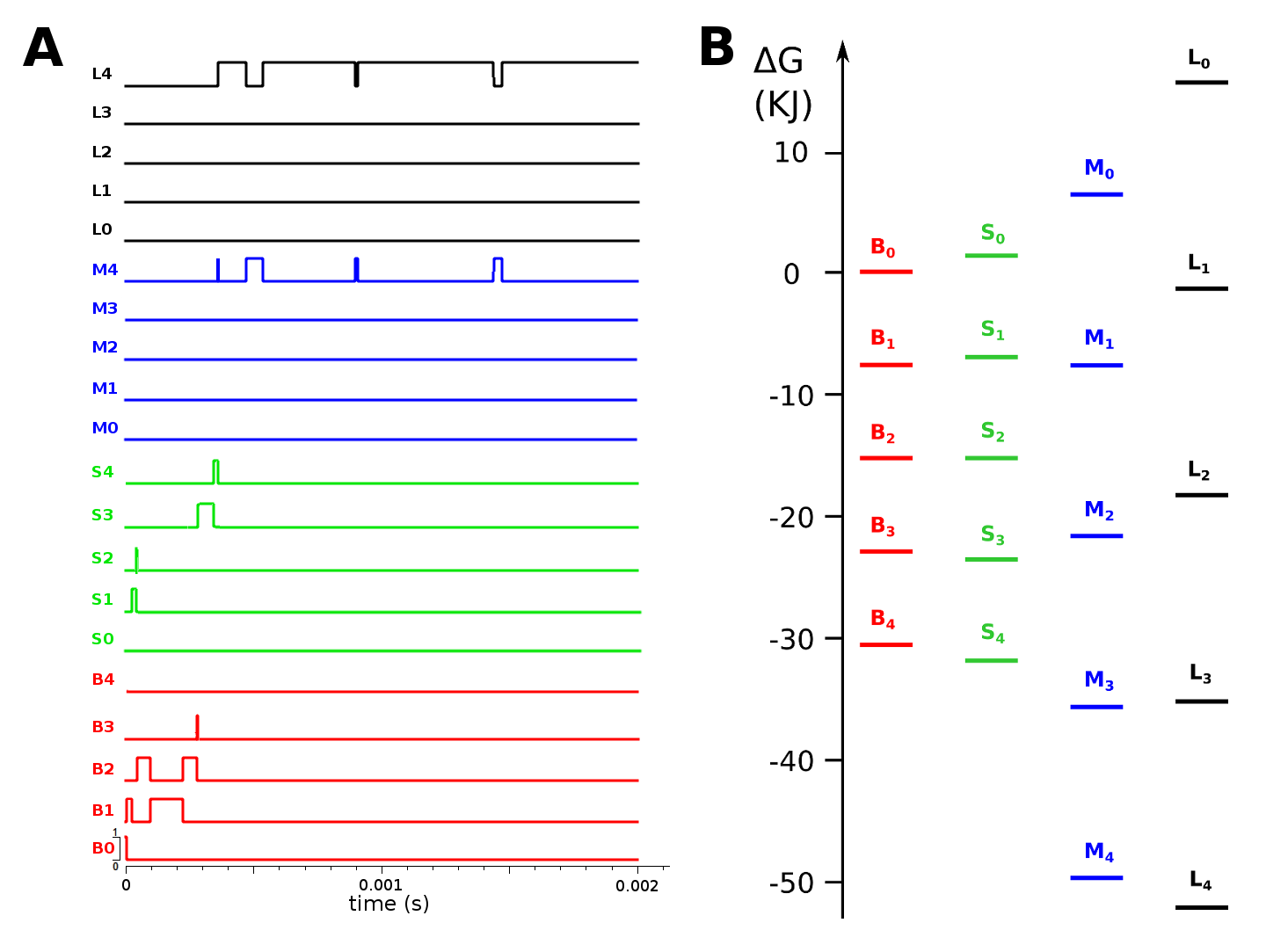}
\caption{\textbf{State transitions of a single channel.} A single channel progresses from a non-liganded basal state to a fully liganded large open state within 0.4 ms upon stimulation with 1~\textmu{}M full agonist (the original STOIC model is provided as Supporting model S4). It should be noted that the simulation is stochastic and this is one of the possible paths the receptor takes to its stable state.}
\label{singlechannel}
\end{center}
\end{figure}

We simulated the time course of a population of AMPA receptors to see how fast the conductances were stabilized upon ligand binding according to predictions of our model, for three different concentrations between 1~\textmu{}M and 1~mM. The higher the starting concentration of the agonist, the faster the population of receptors tends to shift to the large conductance state. With a ligand concentration of 1~mM, the large conductance state is stabilized within the time course, in less than 1~ms as seen in Figure \ref{current} and Figure \ref{singlechannel}. From the kinetic parameters in Table \ref{table3} we can see that ligand affinity increases with the conductance state, but the stabilization of the large conductance state required interconversion rates that were higher than those seen from single molecule FRET data in GluR2 homomeric receptors \cite{Landes2011}. Figure \ref{current} shows that the most stable state was the four-liganded large conductance state, which is also the state the system tended towards in the kinetic simulations as a consequence of it being the state with highest affinity.   
\section*{Discussion}

We have developed a theoretical model in order to formulate a plausible relationship between the conductance states of the AMPA receptor and ligand binding. Our model assumes concerted conformational changes of the receptor subunits, increased ligand affinity with the conductance level, and spontaneous opening of the channel in the absence of ligand. The model can reproduce the thermodynamic behavior of the conductance states upon stimulation with full and partial agonists, as well as the kinetic behavior of a population of receptors in a whole excitatory hippocampal neuron.      

\subsection*{Ligand binding versus ion channel opening}

One of the main predictions of the model is the existence of spontaneous opening of the ion channel. Such events occur at both low and high conductance states, but are less common at high conductance states, as reflected in the allosteric constants in Tables 1 and 2. Structural data which show the direct link between ligand binding and receptor conductance are sparse. However, a full-length GluR2 homomeric receptor has been crystallized bound to an antagonist with the ion pore in a closed conformation \cite{Sobolevsky2009a}. The structure displays a four-fold symmetry in the channel domain. Single-molecule FRET showed that the ligand binding domain could visit four conformational states by each liganded state \cite{Landes2011}. Assuming that the conformational changes in the ligand binding domain are directly linked to conformational changes in the ion channel, this speaks against the theory that each conductance level should correspond to a specific liganded state. Single-channel data also do not tell us how ligands with different efficacy act on the ion pore. The fact that each liganded state can visit different conductance states also supports cooperation of GluR subunits in opening the pore, and evidence for pre-activation states in which the subunits interact have been seen by crosslinking two ligand binding domains \cite{Lau2013}. 

It is important to note the fact that the non-liganded conductance states of ligand-gated ion channels are not only receptor subtype dependent, but they also depend on the expression system and the environment \cite{Changeux1998b}. For instance at low pH, protons can readily affect the conformation of the pores and this has been seen to have a role in spontaneous opening of some channels \cite{Bocquet2009a}. An interesting question would be the physiological advantage of spontaneously open AMPA receptors. This can be of importance for glial cells that are usually acidic and are known to express AMPA receptors \cite{Lu2010a}. The low pH might lead to a higher basal activity of AMPA receptors and thereby contribute to their role in motor activity in these cells \cite{Saab2012a}. Another factor that might increase spontaneous opening of AMPA receptors is electrostatic interactions with surrounding phosholipids in the membrane. This could explain why spontaneous openings are dependent on the expression system. For instance voltage-gated potassium channels, which are evolutionarily related to AMPA receptors's transmembrane domains, exhibit a left-shift in the activation curve in the presence of polyunsaturated fatty acids \cite{Tigerholm2012a}.   

\subsection*{Plausibility of the AMPA receptor gating model}

We chose an MWC-like concerted model to see if we could explain the relationship between ligand binding and the conductance states of the AMPA receptor, since many of its assumptions are fulfilled by existing data. Other allosteric models are based on sequential openings with ligand binding \cite{Koshland1966a, Jonas1993,Diamond1997,Raghavachari2004a}.The MWC theory requires that the protein subunits interconvert between at least two main conformations in a concerted fashion. Hence, the change of conformation is not strictly directed by the binding of a ligand and conformational change of a single ligand binding subunit never occurs, which would likely cause disruption of intersubunit bonds. A consequence is that the state function for the large state does not necessarily overlap with the binding function \cite{Edelstein1997b}. The theory also predicts spontaneously active states. Although originally developed for allosteric enzymes, the theory has been extended to explain behaviour of membrane receptors with several states. It has in particular been used to study the behavior of neurotransmitter receptors such as nicotinic acetylcholine receptor \cite{Edelstein1996b,  Prince1999a, Calimet2013} and of voltage-gated potassium channels \cite{Horrigan2002a} which possess a transmembrane structure similar to AMPAR's. 

It has been hypothesized that the agonist concentration dependence of the three active conductance states visited in native and recombinant AMPA receptors was directly linked to receptor occupancy \cite{Rosenmund1998, Smith2000}. For instance, one interpretation of the GluK2/GluA3flip receptor was that it can only be active if it is bound to at least two agonist molecules \cite{Jin2003, Rosenmund1998}. In this report we show that these single-channel data sets can also be described by concerted conformational changes, which are not directly linked to the liganded state. The existence of active non-liganded states has been identified for other voltage-gated and ligand-gated ion channels \cite{Jackson1984,Turecek1997,Bocquet2009a}. Our model predicts that both recombinant GluK2/GluA3flip receptors expressed in HEK cells stimulated with quisqualate and native receptors expressed in granular cells stimulated with glutamate visit all conductance states in the non-liganded form. However, the frequency of these spontaneous openings decrease with increased conductance. Further, the model predicts that both quisqualate and glutamate stimulation stabilize the large conductance state, as expected for full agonists. In contrast, fitting single-channel data of homomeric GluR2 channels expressed in oocytes that were stimulated with the BrW and IW to the model predicted a stabilization of the large and the medium conductance states respectively, reflecting the ability of partial agonists to stabilize any state. 
      
At this stage, the experimental insights necessary to elucidate the agonist concentration dependence of the conductance states are restricted by the limitations of the experimental measurements. A thorough kinetic analysis of the receptor by locking single channels into each possible liganded state with the use of a photoaffinity analogue to glutamate could be useful in confirming the kinetic framework we have suggested to explain the conductance states \cite{Ruiz1999a}. Another possibility would be to perform single-channel recordings with a mutation in one of the ligand-binding domains. However, since low conductance states are infrequent at highly liganded states this will probably be a time-consuming project.              

Our modeling suggests that describing the behavior of the AMPAR conductance states with a concerted model could improve existing models of synaptic plasticity, and in particular models of the biochemical cascade underlying LTP. In turn this could increase our understanding of the processed underlying learning and memory.

\section*{Materials and Methods}

\subsection*{Preparation, electrophysiology and data analysis of autaptic neurons}
Hippocampal neurons were cultured on astrocyte islands as previously described \cite{Bekkers1991a}. Single-electrode whole-cell voltage-clamp measurements were carried out on embryonic neurons, growing individually on an island between days 12 to 18 after culture. Neurons were voltage-clamped at -70~mV and depolarized to 0~mV for 2~ms to evoke an unclamped action potential. Cells with a response inferior to 500~pA were excluded. For measurements we used an Axopatch 200B amplifier (Molecular Devices) and the Clampex 10.0 software from Molecular Devices. Data acquisition was done at a sampling rate of 10 kHz and low-pass filtered at 3kHz. The series resistance ranged between 7 and 15 MOhm and was compensated for.      

The extracellular medium contained 136~mM NaCl, 2.5~mM KCl, 10~mM glucose, 10~mM HEPES, 2~mM CaCl2, 4~mM MgCl2 and the internal solution contained 140~mM potassium gluconate, 10~mM HEPES, 1~mM EGTA, 4.6~mM MgCl2, 4~mM Na-ATP, 15~mM creatine phosphate, and 50~U/ml phosphocreatine kinase. To identify excitatory neurons 3~mM KA was applied for 2 s.

\subsection*{Simulations}

The state functions were plotted with MatLab, using parameter values obtained with the parameter optimization in COPASI as described below \cite{Hoops2006a}. Timecourse simulations of the kinetic transitions between different conductance substates were done in STOIC \cite{Edelstein1997b}, a program specifically developed for ligand-gated ion channels and COPASI. The data convolution code was written in MatLab. Models and simulations were encoded in SBML \cite{Hucka2003} and SED-ML \cite{Waltemath2011} respectively and submitted to BioModels Database \cite{Lenov2006} with the identifier MODEL1407160000.

\subsection*{The fitting problem}

To determine the optimal parameter values for the thermodynamic model using the available data set \cite{Rosenmund1998}, we formulated the objective function as the sum of squares between the logarithm (base $10$) of the observed frequency of the conductance states and the logarithm of the theoretical value. The purpose of the logarithm function was to give compatible weights to the frequencies of the different states, as the state functions were later plotted in a semilogarithmic scale. In the equations below, G represents the ligand concentration, and each of the equations 1 to 4 give the square of the difference between the logarithm of a specific state function subtracted by the logarithm of the experimental value of the frequency of a state $y_{ij}$ at a certain concentration. The experimental values belong to a matrix $y$, where the rows represent the four different states and the columns represent the different concentrations. To keep the parameters within a biological range, we restricted Ls between $1$ and $10^6$, and Ks between $10^{-10}$ and $10^{-3}$. These constraints gave a wide parameter space as seven parameters were to be determined over large intervals, thus, we tried to avoid using a brute force method. Moreover, the non-linearity of the objective function rendered it unsuitable to minimize with a gradient method as they tend to converge to local minima. A genetic algorithm has thus been chosen to find the global minimum \cite{Wang1997}. We chose the built-in genetic algorithm of Copasi \cite{Hoops2006a}.      

\small

\begin{equation}
B_G=(log(\frac{\frac{1}{L_B}(1+\frac{G}{K_B})^4}{\frac{1}{L_S}(1+\frac{G}{K_S})^4 + \frac{1}{L_M}(1+\frac{G}{K_M})^4 + \frac{1}{L_L}(1+\frac{G}{K_L})^4 + (1+\frac{G}{K_B})^4})-log(y_{41}))^2
\end{equation}

\begin{equation}
S_G=(log(\frac{\frac{1}{L_S}(1+\frac{G}{K_S})^4}{\frac{1}{L_S}(1+\frac{G}{K_S})^4 + \frac{1}{L_M}(1+\frac{G}{K_M})^4 + \frac{1}{L_L}(1+\frac{G}{K_L})^4 + (1+\frac{G}{K_B})^4})-log(y_{31}))^2
\end{equation}

\begin{equation}
M_G=(log(\frac{\frac{1}{L_M}(1+\frac{G}{K_M})^4}{\frac{1}{L_S}(1+\frac{G}{K_S})^4 + \frac{1}{L_M}(1+\frac{G}{K_M})^4 + \frac{1}{L_L}(1+\frac{G}{K_L})^4 + (1+\frac{G}{K_B})^4})-log(y_{21}))^2
\end{equation}

\begin{equation}
L_G=(log(\frac{\frac{1}{L_L}(1+\frac{G}{K_L})^4}{\frac{1}{L_S}(1+\frac{G}{K_S})^4 + \frac{1}{L_M}(1+\frac{G}{K_M})^4 + \frac{1}{L_L}(1+\frac{G}{K_L})^4 + (1+\frac{G}{K_B})^4})-log(y_{11}))^2
\end{equation}

\begin{equation}
Objective function=\sum_{i=1}^c L_G + \sum_{i=1}^c M_G + \sum_{i=1}^c S_G + \sum_{i=1}^c B_G
\end{equation} 

\normalsize

\subsection*{Inference of kinetic parameters from thermodynamic parameters}
$$L_S=\frac{[B_0]}{[A^S_0]}=\frac{^{SB}k}{^{BS}k}$$
  
$$L_M=\frac{[B_0]}{[A^M_0]}=L_S \times \frac{^{MS}k}{^{SM}k}$$

$$L_L=\frac{[B_0]}{[A^L_0]}=L_S \times \frac{^{MS}k}{^{SM}k} \times \frac{^{LM}k}{^{ML}k}$$

$$K_B=\frac{^Bk_\mathit{off}}{^Bk_{on}}$$

$$K_S=\frac{^Sk_\mathit{off}}{^Sk_{on}}$$

$$K_M=\frac{^Mk_\mathit{off}}{^Mk_{on}}$$

$$K_L=\frac{^Lk_\mathit{off}}{^Lk_{on}}$$

\section*{Acknowledgments}
We would like to thank Lukas Endler, James Lu and Melanie Stefan for advice on parameter optimization. Our gratitude also goes to thesis committee members at Karolinska Institutet for valuable discussions and tutors at the single-channel summer school at University College London.

\bibliographystyle{plos2009}
\bibliography{AMPARbib}

\newpage

\section*{Figure Legends}

\clearpage

\section*{Tables}

\begin{table}[!ht]
\centering
\begin{tabular}{c c c c}
\hline\hline
Parameter & Native & GluA3/GluK2 \\
\hline
$^{BS}c$ & $-$ & $0.0601$ \\		                  
$^{SM}c$ & $0.37$ & $0.0616$ \\	                  
$^{ML}c$ &  $0.2703$ & $0.0690$ \\  
$L_S$ & $20.387$ & $2.867$ \\		                  
$L_M$ & $28.898$ & $207.5$ \\	                  
$L_L$ &  $1066.73$ & $10^6$ \\                     
$K_B$ &  $-$ & $9 \times 10^{-4}$\\                 
$K_S$ & $1 \times 10^{-3}$ & $5.407 \times 10^{-5}$ \\                  
$K_M$ &  $3.7 \times 10^{-4} $ & $3.330 \times 10^{-6}$ \\ 
$K_L$ &  $1 \times 10^{-4}$ & $2.298 \times 10^{-7}$ \\                                      	
\hline\hline
\end{tabular}
\caption{Allosteric and dissociation constants determined for full agonists.}
\label{table1}
\end{table}

\clearpage

\begin{table}[!ht]
\centering
\begin{tabular}{c c c }
\hline\hline
Parameter &  IW & BrW \\
\hline                     
$K_S$ &  $1 \times 10^{-3}$ & $4 \times 10^{-3}$ \\                 
$K_M$ & $8.580 \times 10^{-5}$ & $4.97 \times 10^{-5}$ \\                  
$K_L$ &  $1.198 \times 10^{-5}$ & $5.02 \times 10^{-6}$ \\                                    	
\hline\hline
\end{tabular}
\caption{Dissociation constants for partial agonists.}
\label{table2}
\end{table}

\clearpage

\begin{table}[!ht]
\centering
\begin{tabular}{c c}
\hline\hline
Parameter & Values \\
\hline
$^Bk_\mathit{on}$ & $5 \times 10^6$ M\textsuperscript{-1}.s\textsuperscript{-1} \\		                  
$^Bk_\mathit{off}$ & $4.495 \times 10^3$ s\textsuperscript{-1} \\	                  
$^Sk_\mathit{on}$ &  $5 \times 10^6$ M\textsuperscript{-1}.s\textsuperscript{-1}\\                     
$^Sk_\mathit{off}$ &  $2.704 \times 10^2$ s\textsuperscript{-1}\\                 
$^Mk_\mathit{on}$ & $5 \times 10^6$ M\textsuperscript{-1}.s\textsuperscript{-1}\\                  
$^Mk_\mathit{off}$ &  $16.65$ s\textsuperscript{-1}\\ 
$^Lk_\mathit{on}$ &  $5 \times 10^6$ M\textsuperscript{-1}.s\textsuperscript{-1}\\
$^Lk_\mathit{off}$ &  $1.149$ s\textsuperscript{-1}\\ 
$^{BS}k_4$ & $3 \times 10^5$ s\textsuperscript{-1}\\		                  
$^{SB}k_4$ & $860$ s\textsuperscript{-1}\\	                  
$^{SM}k_4$ &  $20000$ s\textsuperscript{-1}\\                     
$^{MS}k_4$ &  $145$ s\textsuperscript{-1}\\                 
$^{ML}k_4$ & $50000$ s\textsuperscript{-1}\\                  
$^{LM}k_4$ &  $691$ s\textsuperscript{-1}\\
 \hline\hline
\end{tabular}
\caption{Kinetic parameters used for the simulations in figure \ref{current}. The interconversion rates correspond to the fully-liganded state. The free energy gained by the binding of a ligand is assumed to be used to speed-up forward rates and slow down reverse rates. Therefore the forward rate with n ligands bound are obtained by multiplying the rate with n+1 ligands bound by the respective c constant (see Table 1) to the power of the transition parameter. The reverse rates are obtained by dividing with the c constant to the power of the transition parameter. We assumed an even distribution of free energy and used a transition parameter of 0.5 for all transitions. See \cite{Edelstein1996b} for an example of non even distribution.}
\label{table3}
\end{table}

\section*{Supplementary equations}

\subsection*{Definitions}

\begin{equation*}                                                                                               
B=B_0+\frac{4B_0X}{K_B}+\frac{6B_0X^2}{K_B^2}+\frac{4B_0X^3}{K_B^3}+\frac{B_0X^4}{K_B^4}
\end{equation*}

\begin{equation*}
S=A_0^S+\frac{4A_0^SX}{K_S}+\frac{6A_0^SX^2}{K_s^2}+\frac{4A_0^SX^3}{K_s^3}+\frac{A_0^SX^4}{K_s^4}
\end{equation*}

\begin{equation*}
M=A_0^M+\frac{4A_0^MX}{K_M}+\frac{6A_0^MX^2}{K_M^2}+\frac{4A_0^MX^3}{K_M^3}+\frac{A_0^MX^4}{K_M^4}
\end{equation*}

\begin{equation*}
L=A_o^L+\frac{4A_0^LX}{K_L}+\frac{6A_0^LX^2}{K_L^2}+\frac{4A_0^LX^3}{K_L^3}+\frac{A_0^LX^4}{K_L^4}
\end{equation*}

\begin{equation*}
\frac{B}{A_0^L}=L_SL_ML_L+\frac{4L_SL_ML_LX}{K_B}+\frac{6L_SL_ML_LX^2}{K_B^2}+\frac{4L_SL_ML_LX^3}{K_B^3}+\frac{L_SL_ML_LX^4}{K_B^4}
\end{equation*}

\begin{equation*}
\frac{S}{A_0^L}=L_ML_L+\frac{4L_ML_LX}{K_S}+\frac{6L_ML_LX^2}{K_S^2}+\frac{4L_ML_LX^3}{K_S^3}+\frac{L_ML_LX^4}{K_S^4}
\end{equation*}

\begin{equation*}
\frac{M}{A_0^L}=L_L+\frac{4L_LX^2}{K_M}+\frac{6L_LX^2}{K_M^2}+\frac{4L_LX^3}{K_M^3}+\frac{L_LA_0^LX^4}{K_M^4}
\end{equation*}

\begin{equation*}
\frac{L}{A_0^L}=1+\frac{4X}{K_L}+\frac{6X^2}{K_L^2}+\frac{4X^3}{K_L^3}+\frac{X^4}{K_L^4}
\end{equation*}

\subsection*{Derivation of the saturation function}

\tiny

\begin{equation*}
\bar{Y}=\frac{B_1 + 2B_2 + 3B_3 + 4B_4 + A_1^S + 2A_2^S + 3A_3^S + 4A_4^S + A_1^M + 2A_2^M + 3A_3^M + 4A_4^M + A_1^L + 2A_2^L + 3A_3^L + 4A_4^L}{4(B_0 + B_1 + B_2 + B_3 + B_4 + A_0^S + A_1^S + A_2^S + A_3^S + A_4^S + A_0^M + A_1^M + A_2^M + A_3^M + A_4^M + A_0^L + A_1^L + A_2^L + A_3^L + A_4^L )}
\end{equation*}

\begin{equation*}
\bar{Y}=\frac{B_0\frac{X}{K_B}(1+\frac{X}{K_B})^3 + A_0^S\frac{X}{K_S}(1+\frac{X}{K_S})^3 + A_0^M\frac{X}{K_M}(1+\frac{X}{K_M})^3 + A_0^L\frac{X}{K_L}(1+\frac{X}{K_L})^3}{B_0\frac{X}{K_B}(1+\frac{X}{K_B})^4+A_0^S\frac{X}{K_S}(1+\frac{X}{K_S})^3 + A_0^M\frac{X}{K_M}(1+\frac{X}{K_M})^3 + A_0^L\frac{X}{K_L}(1+\frac{X}{K_L})^3}
\end{equation*}

\begin{equation*}
\bar{Y}=\frac{B_0\frac{X}{K_B}(1+\frac{X}{K_B})^3 + \frac{B_0}{L_S}\frac{X}{K_S}(1+\frac{X}{K_S})^3 + \frac{B_0}{L_M}\frac{X}{K_M}(1+\frac{X}{K_M})^3 + \frac{B_0}{L_L}\frac{X}{K_L}(1+\frac{X}{K_L})^3}{B_0\frac{X}{K_B}(1+\frac{X}{K_B})^4+\frac{B_0}{L_S}\frac{X}{K_S}(1+\frac{X}{K_S})^4 + \frac{B_0}{L_M}\frac{X}{K_M}(1+\frac{X}{K_M})^4 + \frac{B_0}{L_L}\frac{X}{K_L}(1+\frac{X}{K_L})^4}
\end{equation*}

\begin{equation*}
Y=\frac{B_0\frac{X}{K_B}(1+\frac{X}{K_B})^3 + \frac{1}{L_S}\frac{X}{K_S}(1+\frac{X}{K_S})^3 + \frac{1}{L_M}\frac{X}{K_M}(1+\frac{X}{K_M})^3 + \frac{1}{L_L}\frac{X}{K_L}(1+\frac{X}{K_L})^3}{\frac{X}{K_B}(1+\frac{X}{K_B})^4+\frac{1}{L_S}\frac{X}{K_S}(1+\frac{X}{K_S})^4 + \frac{1}{L_M}\frac{X}{K_M}(1+\frac{X}{K_M})^4 + \frac{1}{L_L}\frac{X}{K_L}(1+\frac{X}{K_L})^4}
\end{equation*}

\normalsize

\subsection*{Derivation of a state function, example of the large conductance}

\tiny

\begin{equation*}
A_L=\frac{A_o^L+A_1^L+A_2^L+A_3^L+A_4^L}{B_0+B_1+B_2+B_3+B_4+A_o^s+A_1^L+A_2^s+A_3^s+A_4^s+A_o^M+A_1^M+A_2^M+A_3^M+A_4^M+A_o^L+A_1^L+A_2^L+A_3^L+A_4^L}
\end{equation*}


\begin{equation*}
A_L=\frac{A_o^L+\frac{4A_0^LX}{K_L}+\frac{6A_0^LX^2}{K_L^2}+\frac{4A_0^LX^3}{K_L^3}+\frac{A_0^LX^4}{K_L^4}}{B+S+M+L}
\end{equation*}

\begin{equation*}
A_L=\frac{1+\frac{4X}{K_L}+\frac{6X^2}{K_L^2}+\frac{4X^3}{K_L^3}+\frac{X^4}{K_L^4}}{(B+S+M+L)/A_0^L}
\end{equation*}


\begin{equation*}
A_L=\frac{(1+\frac{X}{K_L})^4}{L_SL_ML_L(1+\frac{X}{K_B})^4+L_ML_L(1+\frac{X}{K_S})^4+L_L(1+\frac{X}{K_M})^4+(1+\frac{X}{K_L})^4}
\end{equation*}

\normalsize

\subsection*{Kinetic model}

\tiny

\begin{equation*}
\frac{d[B_0]}{dt}=-4^Bk_\mathit{on}[B_0][G] + ^Bk_\mathit{off}[B_1] - ^{BS}k_0[B_0] + ^{SB}k_0[A_0^S]
\end{equation*}

\begin{equation*}
\frac{d[B_1]}{dt}=-3^Bk_\mathit{on}[B_1][G] + 2^Bk_\mathit{off}[B_2] - ^Bk_\mathit{off}[B_1] + 4^Bk_\mathit{on}[B_0][G] - ^{BS}k_1[B_1]+ ^{SB}k_1[A_1^S]
\end{equation*}

\begin{equation*}
\frac{d[B_2]}{dt}=-2^Bk_\mathit{on}[B_2][G] + 3^Bk_\mathit{off}[B_3] - 2^Bk_\mathit{off}[B_2] + 3^Bk_\mathit{on}[B_1][G] - ^{BS}k_2[B_2]+ ^{SB}k_2[A_2^S]
\end{equation*}

\begin{equation*}
\frac{d[B_3]}{dt}=-^Bk_\mathit{on}[B_3][G] + 4^Bk_\mathit{off}[B_4] - 3^Bk_\mathit{off}[B_3] + 2^Bk_\mathit{on}[B_2][G]- ^{BS}k_3[B_3]+ ^{SB}k_3[A_3^S]
\end{equation*}

\begin{equation*}
\frac{d[B_4]}{dt}=- 4^Bk_\mathit{off}[B_4] + ^Bk_\mathit{on}[B_3][G] - ^{BS}k_4[B_4] + ^{SB}k_3[A_4^S]
\end{equation*}

\begin{equation*}
\frac{d[A^S_0]}{dt}=-4^Sk_\mathit{on}[A_0^S][G] + ^Sk_\mathit{off}[A_1^S]                                                         - ^{SM}k_0[A_0^S] + ^{MS}k_0[A_0^M] - ^{SB}k_0[A_0^S] + ^{BS}k_0[B_0]
\end{equation*}

\begin{equation*}
\frac{d[A^S_1]}{dt}=-3^Sk_\mathit{on}[A_1^S][G] + 2^Sk_\mathit{off}[A_2^S] - ^Sk_\mathit{off}[A_1^S] + 4^Sk_\mathit{on}[A_0^S][G] - ^{SM}k_1[A_1^S] + ^{MS}k_1[A_1^M] - ^{SB}k_1[A_1^M] + ^{BS}k_1[B_1]
\end{equation*}

\begin{equation*}
\frac{d[A^S_2]}{dt}=-2^Sk_\mathit{on}[A_2^S][G] + 3^Sk_\mathit{off}[A_3^S] - 2^Sk_\mathit{off}[A_2^S] + 3^Sk_\mathit{on}[A_1^S][G] - ^{SM}k_2[A_2^S] + ^{MS}k_2[A_2^M] - ^{SB}k_2[A_2^M] + ^{BS}k_2[B_2]
\end{equation*}

\begin{equation*}
\frac{d[A^S_3]}{dt}=-^Sk_\mathit{on}[A_3^S][G] + 4^Sk_\mathit{off}[A_4^S] - 3^Sk_\mathit{off}[A_3^S] + 2^Sk_\mathit{on}[A_2^S][G] - ^{SM}k_3[A_3^S] + ^{MS}k_3[A_3^M] - ^{SB}k_3[A_3^M] + ^{BS}k_3[B_3]
\end{equation*}

\begin{equation*}
\frac{d[A^S_4]}{dt}=-4^Sk_\mathit{off}[A_4^S] + ^Sk_\mathit{on}[A_3^S][G]                                                         - ^{SM}k_4[A_4^S] + ^{MS}k_4[A_4^M] - ^{SB}k_4[A_4^M] + ^{BS}k_4[B_4]
\end{equation*}

\begin{equation*}
\frac{d[A^M_0]}{dt}=-4^Mk_\mathit{on}[A_0^M][G] + ^Mk_\mathit{off}[A_1^M]                                                         - ^{ML}k_0[A_0^M] + ^{LM}k_0[A_0^L] - ^{MS}k_0[A_0^M] + ^{SM}k_0[A_0^S]
\end{equation*}

\begin{equation*}
\frac{d[A^M_1]}{dt}=-3^Mk_\mathit{on}[A_1^M][G] + 2^Mk_\mathit{off}[A_2^M] - ^Mk_\mathit{off}[A_1^M] + 4^Mk_\mathit{on}[A_0^M][G] - ^{ML}k_1[A_1^M] + ^{LM}k_1[A_1^L] - ^{MS}k_1[A_1^M] + ^{SM}k_1[A_1^S]
\end{equation*}

\begin{equation*}
\frac{d[A^M_2]}{dt}=-2^Mk_\mathit{on}[A_2^M][G] + 3^Mk_\mathit{off}[A_3^M] - 2^Mk_\mathit{off}[A_2^M] + 3^Mk_\mathit{on}[A_1^M][G] - ^{ML}k_2[A_2^M] + ^{LM}k_2[A_2^L] - ^{MS}k_2[A_2^M] + ^{SM}k_2[A_2^S]
\end{equation*}

\begin{equation*}
\frac{d[A^M_3]}{dt}=-^Mk_\mathit{on}[A_3^M][G] + 4^Mk_\mathit{off}[A_4^M] - 3^Mk_\mathit{off}[A_3^M] + 2^Mk_\mathit{on}[A_2^M][G] - ^{ML}k_3[A_3^M] + ^{LM}k_3[A_3^L] - ^{MS}k_3[A_3^M] + ^{SM}k_3[A_3^S]
\end{equation*}

\begin{equation*}
\frac{d[A^M_4]}{dt}=-4^Mk_\mathit{off}[A_4^M] + ^Mk_\mathit{on}[A_3^M][G]                                                         - ^{ML}k_4[A_4^M] + ^{LM}k_4[A_4^L] - ^{MS}k_4[A_4^M] + ^{SM}k_4[A_4^S]
\end{equation*}

\begin{equation*}
\frac{d[A^L_0]}{dt}=-4^Lk_\mathit{on}[A_0^L][G] + ^Lk_\mathit{off}[A_1^L]                                                         - ^{LM}k_0[A_0^L] + ^{ML}k_0[A_0^M]
\end{equation*}

\begin{equation*}
\frac{d[A^L_1]}{dt}=-3^Lk_\mathit{on}[A_1^L][G] + 2^Lk_\mathit{off}[A_2^L] - ^Lk_\mathit{off}[A_1^L] + 4^Lk_\mathit{on}[A_0^L][G] - ^{LM}k_1[A_1^L] + ^{ML}k_1[A_1^M]
\end{equation*}

\begin{equation*}
\frac{d[A^L_2]}{dt}=-2^Lk_\mathit{on}[A_2^L][G] + 3^Lk_\mathit{off}[A_3^L] - 2^Lk_\mathit{off}[A_2^L] + 3^Lk_\mathit{on}[A_1^L][G] - ^{LM}k_2[A_2^L] + ^{ML}k_2[A_2^M]
\end{equation*}

\begin{equation*}
\frac{d[A^L_3]}{dt}=-^Lk_\mathit{on}[A_3^L][G] + 4^Lk_\mathit{off}[A_4^L] - 3^Lk_\mathit{off}[A_3^L] + 2^Lk_\mathit{on}[A_2^L][G] - ^{LM}k_3[A_3^L] + ^{ML}k_3[A_3^M]
\end{equation*}

\begin{equation*}
\frac{d[A^L_4]}{dt}=-4^Lk_\mathit{off}[A_4^L] + ^Lk_\mathit{on}[A_3^L][G]                                                         - ^{LM}k_4[A_4^L] + ^{ML}k_4[A_4^M]
\end{equation*}

\end{document}